\documentclass[aps,prb,reprint]{revtex4-1}
\usepackage{blindtext}
\usepackage{graphicx}
\usepackage{color}
\usepackage{amsbsy}
\begin{document}
\title{Tuning the thermal conductance of molecular junctions with interference effects}
\author{J. C. Kl\"{o}ckner$^{1}$}
\email{Jan.Kloeckner@uni-konstanz.de}
\author{J. C. Cuevas$^{1,2}$}
\author{F. Pauly$^{1,3}$}

\affiliation{$^{1}$Department of Physics, University of Konstanz, D-78457
  Konstanz, Germany} 
\affiliation{$^{2}$Departamento de F\'{\i}sica Te\'orica
  de la Materia Condensada and Condensed Matter Physics Center (IFIMAC),
  Universidad Aut\'onoma de Madrid, E-28049 Madrid, Spain}
\affiliation{$^{3}$Okinawa Institute of Science and Technology Graduate
  University, Onna-son, Okinawa 904-0395, Japan}

\begin{abstract}
We present an \emph{ab initio} study of the role of interference effects in
the thermal conductance of single-molecule junctions. To be precise, using a
first-principles transport method based on density functional theory, we
analyze the coherent phonon transport in single-molecule junctions made of
several benzene and oligo-phenylene-ethynylene derivatives. We show that the
thermal conductance of these junctions can be tuned via the inclusion of
substituents, which induces destructive interference effects and results in a
decrease of the thermal conductance with respect to the unmodified
molecules. In particular, we demonstrate that these interference effects
manifest as antiresonances in the phonon transmission, whose energy positions
can be tuned by varying the mass of the substituents. Our work provides
clear strategies for the heat management in molecular junctions and more
generally in nanostructured metal-organic hybrid systems, which are important
to determine, how these systems can function as efficient energy-conversion
devices such as thermoelectric generators and refrigerators.
\end{abstract}

\maketitle

\section{Introduction}
The manipulation of phonon heat conduction at the nanoscale is of fundamental
interest for technologies such as thermoelectrics and thermal management in
nanoelectronics \cite{Cahill2014}. Earlier attempts to perform such a
manipulation focused on incoherent transport mechanisms, but in recent years,
and with the advent of nanoscale devices and nanostructured materials, a great
effort is being devoted to control heat currents in the coherent regime by
making use of interference effects \cite{Maldovan2015}. Thus for instance, in
the context of phononic structures the so-called superlattices have been
introduced. Here one can tune the phonon band structure, phonon group
velocity, and the related phonon density of states by means of wave
interference effects \cite{Luckyanova2012,Ravichandran2014}. This strategy is
limited by the quality of interfaces in terms of interface roughness and by
the mean free path of the heat carriers. Due to these constraints the strategy
is in general better applicable to systems, in which heat is carried by
low-energy or equivalently long-wavelength phonons. However, if one assumes
that atomically thin, single crystal planes can be manipulated at will, also
higher-energy or short-wavelength phonons can in principle be affected, since
structural length scales are of interatomic distance. Thus, Han \emph{et
  al.}~\cite{Han2014} theoretically proposed the suppression of heat transport
using defect atom arrays embedded in a single crystal plane. To inhibit the
heat flow, the authors exploited the destructive interference between two
phonon paths.  Resulting line shapes of the energy-dependent phonon
transmission are indeed reminiscent of Fano resonances. Such Fano resonances
are, in fact, a very general concept in nanostructured materials and occur in
every system, in which a narrow discrete mode couples to a broad continuous
spectrum \cite{Fano1961}. Indeed, they have been reported in a great variety
of systems ranging from quantum dots to photonic structures
\cite{Miroshnichenko2010,Lukyanchuk2010,Fan2014}.

As compared to the inorganic materials discussed above, molecules offer an
ideal platform to tailor structures at the single-atom level, and
single-molecule junctions can be used to probe the coherent transport through
the molecules \cite{Cuevas2017}. In the field of molecular electronics
interference effects have been studied with a special emphasis on their impact
on the electronic transport properties of such junctions.  Theoretically this
topic has been explored extensively in the last two decades
\cite{Sautet1988,Emberly1998,Baer2002,Stadler2003,Walter2004,Cardamone2006,Stafford2007,
  Ke2008,Andrews2008a,Solomon2008a,Solomon2008b,Andrews2008b,Stadler2009,Solomon2010,
  Markussen2010a,Geranton2013,Zotti2013}. In particular, special attention has
been devoted to the role of Fano resonances
\cite{Grigoriev2006,Papadopoulos2006,Ernzerhof2007,Shi2007,Finch2009} and to
the determination of general rules governing the appearance of quantum
interference effects in molecules with extended $\pi$-electron systems
\cite{Yoshizawa2008,Markussen2010b,Pedersen2014,Garner2016}. Also different
experimental reports in recent years have convincingly shown the influence of
the quantum interference on electronic transport of molecular junctions, as
illustrated by measurements of linear conductances and current-voltage
characteristics
\cite{Mayor2003,Fracasso2011,Hong2012,Guedon2012,Aradhya2012,Vazquez2012,Rabache2013,
  Arroyo2013a,Arroyo2013b,Xia2014,Manrique2015,Frisenda2016}. When comparing
electron and phonon transport, it needs to be kept in mind that for electrons
the interference has to occur within some $k_{\text{B}}T$ around $\mu\approx
E_{\text{F}}$, with the Fermi energy $E_{\text{F}}$, to get a measurable
effect on the linear conductance. For phonons instead, Debye energies of
typical metal electrodes are in the range of several ten meV so that already a
sizable window of phonon energies contributes to thermal transport at room
temperature.

In the case of coherent phonon transport in molecular junctions the impact of
interference effects is starting to be analyzed theoretically
\cite{Markussen2013,Famili2017}. For instance, Markussen \cite{Markussen2013}
investigated the role of phonon interference in molecular junctions made of
benzene and oligo-phenylene-ethynylene (OPE3) molecules attached to Si and
graphene nanoribbon electrodes. Combining \emph{ab initio} calculations for
vibrational properties of the molecules with a phenomenological description of
the leads and the molecule-lead couplings, Markussen found that the phonon
transmission function for cross-conjugated molecules, like meta-connected
benzene, exhibits destructive interference features very similar to those
found for the corresponding electronic transport, which cause a reduction of
the phononic thermal conductance with respect to the linearly conjugated
analogues. On the other hand, Famili \emph{et al.} \cite{Famili2017} studied the
phonon transport in alkane chains by means of a first-principles method based
on density functional theory (DFT). In particular, they investigated the
appearance of Fano resonances, when the alkanes are modified by the inclusion
of certain side groups, so-called ``Christmas trees''.  These resonances led
to a reduction of the corresponding thermal conductance by a factor
2. However, alkanes are known to exhibit geometrical gauche defects that
result in the localization of vibrational modes. These defects have been
reported to reduce the thermal conductance by a similar magnitude
\cite{Li2015}, making it presumably difficult to discriminate between the
effects of side-groups and gauche defects. In this sense, the study of stiff
molecules like benzene derivatives may provide more conclusive results about
the existence of interference effects.

In this work we employ a full \emph{ab initio} DFT-based transport method to
study the role of phonon interference effects in the thermal conductance of
single-molecule junctions. In particular we explore benzene- and OPE3-related
molecules, but contrary to Ref.~[\onlinecite{Markussen2013}] we assume that the
electrodes are made of Au, which is often the material of choice for the leads
of molecular junctions, and use amine anchoring groups. In the case of
benzenediamine we find that due to the small Debye energy of Au of around 20
meV, no destructive interference effects are visible in the phonon
transmission function, irrespective of whether the molecule is contacted to Au
in para or meta configuration.  This leads to a room-temperature thermal
conductance that is similar for the two contacting schemes. More importantly
we show that this situation can be changed by replacing a H atom of benzene by
a halogen atom (F, Cl, Br, I). The substitution may lead to a reduction of the
thermal conductance by up to a factor of 1.7. We also show that by increasing
the number of substituent atoms in the benzene molecule and depending on their
precise position on the ring, one can induce additional reductions of the
thermal conductance by a factor of 2.5. Finally, we also show that similar
concepts apply to the case of OPE3 and, in particular, we find a clear
difference between para- and meta-OPE3, where the central benzene ring is
connected in para or meta position. Our work provides concrete predictions
that can potentially be tested, given the recent experimental advances in the
measurement of the thermal conductance of atomic-scale contacts
\cite{Longi2017a,Mosso2017b}. In addition, it sheds light on the importance of
phonon interference effects to tune the thermal transport of molecular
junctions.

The rest of the paper is organized as follows. In section II we briefly
describe the theoretical techniques employed in this work to study the phonon
transport in single-molecule junctions. In section III we present the main
results of this work concerning the phonon thermal conductance of
single-molecule junctions based on benzene and OPE3 derivatives. We summarize
our main conclusions in section IV. Finally, we discuss in
  Appendix \ref{sec-elthcond} our results for the electronic thermal conductance of the
  molecular junctions studied in this work to show that the thermal transport
  is actually dominated by phonons.

\section{Theoretical method}

To explore the influence of phonon interference on the thermal conductance of
single-molecule junctions, we describe coherent phonon transport within the
Landauer-B\"uttiker approach. This approach is based on the
  harmonic approximation and is valid, if the characteristic dimensions of the
  atomic-scale junction are smaller than the inelastic mean free path for
  phonons, which is on the order of a few nm at room temperature for gold
  \cite{Jain2016}. The main source of resistance is then the elastic
  scattering in the narrowest part of the device. Thermalization due to third
  or higher order interactions between atoms that lead to phonon-phonon
  scattering is assumed to take place exclusively in the electrodes, which are
  thus treated as reservoirs with a well-defined thermodynamic state.

Within this approach the linear thermal conductance due to phonons is given by
\begin{equation}
  \kappa_{\rm pn}(T) = \frac{1}{h} \int_{0}^{\infty} dE \,E \tau_{\rm pn}(E)
  \frac{\partial n(E,T)}{\partial T}, \label{eq-kph}
\end{equation}
where $n(E,T)=[\exp(E/k_{\rm B}T)-1]^{-1}$ is the Bose function and $\tau_{\rm
  pn}(E)$ is the energy-dependent phononic transmission. We compute the
transmission function $\tau_{\rm pn}(E)$ by means of a combination of DFT and
non-equilibrium Green's function techniques, as we have described in detail in
Refs.~[\onlinecite{Buerkle2015,Kloeckner2016,Kloeckner2017}].

Briefly, the first step in the calculation is the construction of the
molecular junction geometries. For this purpose, we use DFT to obtain
equilibrium geometries through total energy minimization. From these
calculations the vibrational properties in terms of the dynamical matrix are
obtained by applying density functional perturbation theory, as implemented in
the quantum chemistry software package TURBOMOLE 6.5
\cite{TURBOMOLE,Deglmann2002,Deglmann2004}. In our DFT calculations we employ
the Perdew-Burke-Ernzerhof exchange-correlation functional
\cite{Perdew1992,Perdew1996}, the basis set of split-valence-plus-polarization
quality def2-SV(P) \cite{Weigend2005}, and the corresponding Coulomb fitting
basis \cite{Weigend2006}. In order to accurately determine the force constants
and related vibrational energies, we use very strict convergence criteria. In
particular, total energies are converged to a precision of better than
$10^{-9}$~a.u., whereas geometry optimizations are performed until the change
of the maximum norm of the Cartesian gradient is below $10^{-5}$~a.u.  We have
checked that our stringent convergence criteria avoid the appearance of any
modes with imaginary frequencies in the optimized junction region, which would
otherwise signal unstable geometries. It is furthermore worth stressing that
in our method the electrodes are described by means of perfect semi-infinite
crystals, whose phonon properties are determined within DFT with the same
functional and the same basis set as used for the central device part. In this
way we achieve a consistent, full \emph{ab initio} treatment of the phonon
system of the whole molecular junctions. Finally, the dynamical matrix of the
molecular junction is used to compute the phonon transmission function with
the help of non-equilibrium Green's function techniques, as presented in
Ref.~[\onlinecite{Buerkle2015}].

\section{Results}

\begin{figure}[t]
\centering
\includegraphics[width=0.4\textwidth]{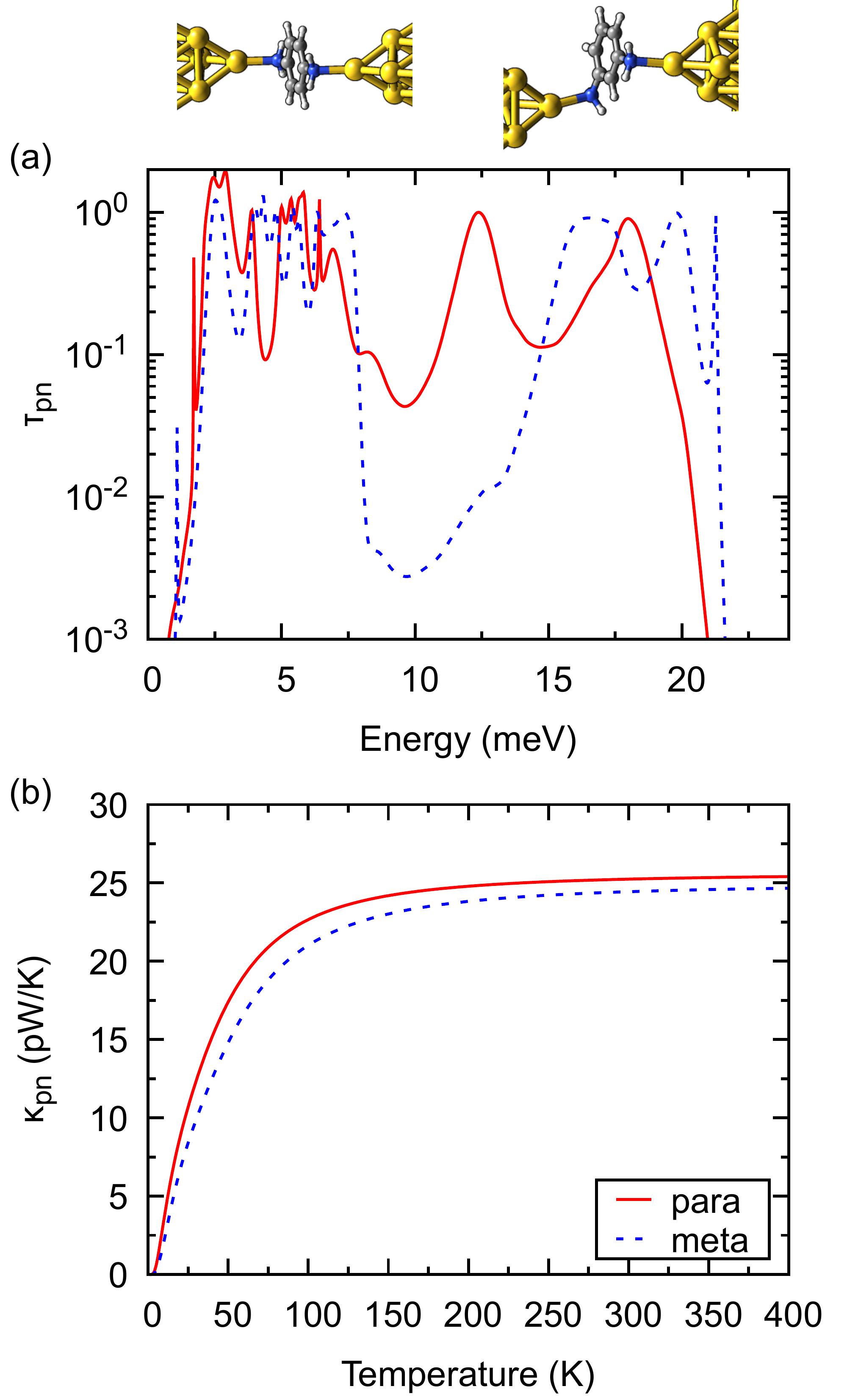}
\caption{(a) Phonon transmission as a function of energy for the
  Au-benzenediamine-Au junctions shown above this panel for
  both para (left) and meta (right) binding configurations. (b) The
  corresponding phononic thermal conductance as a function of
  temperature. The legend in panel (b) applies also to panel
    (a).}
\label{fig-Benzene-para-meta}
\end{figure}

We start our discussion of the results with an analysis of single-molecule
junctions based on the unsubstituted benzenediamine molecule. As shown in the
upper part of Fig.~\ref{fig-Benzene-para-meta}, we consider contacts, where
the amino (NH$_2$) group is attached to a single tip atom of the gold
electrodes on each side both in the para and in the meta configuration. We
will refer to this amino binding site on the gold also as ``atop position''
and note that these geometries are similar to those used in previous studies
of electronic transport, mimicking typical binding geometries
\cite{Quek2007}. In Fig.~\ref{fig-Benzene-para-meta}(a) we show the results
for the phononic transmission of these two binding configurations, computed
with our \emph{ab initio} method, described in the previous section. The first
thing to notice is that the transmission is only finite below approximately 20
meV, which corresponds to the Debye energy of the gold electrode material. On
the other hand, notice that although both transmission curves are different,
which is reasonable due to the different geometrical configurations, we do not
find any signature of destructive interference in the form of
antiresonances. This is further confirmed by the results for the temperature
dependence of the phononic thermal conductance, which we show in
Fig.~\ref{fig-Benzene-para-meta}(b). In fact, both molecules exhibit similar
thermal conductance values over the whole temperature range explored here.
 
The reason for the lack of destructive interference effects in these
benzene-based junctions can be understood with the help of the work of
Markussen \cite{Markussen2013}. Considering his semi-empirical results for
phonon transport in benzene junctions with Si electrodes, he found that the
lowest observable destructive interference features appear at energies around
40 meV. Since this energy is above the Debye energy of gold, no effects are
visible in our case.

\begin{figure}[t]
\centering
\includegraphics[width=0.4\textwidth]{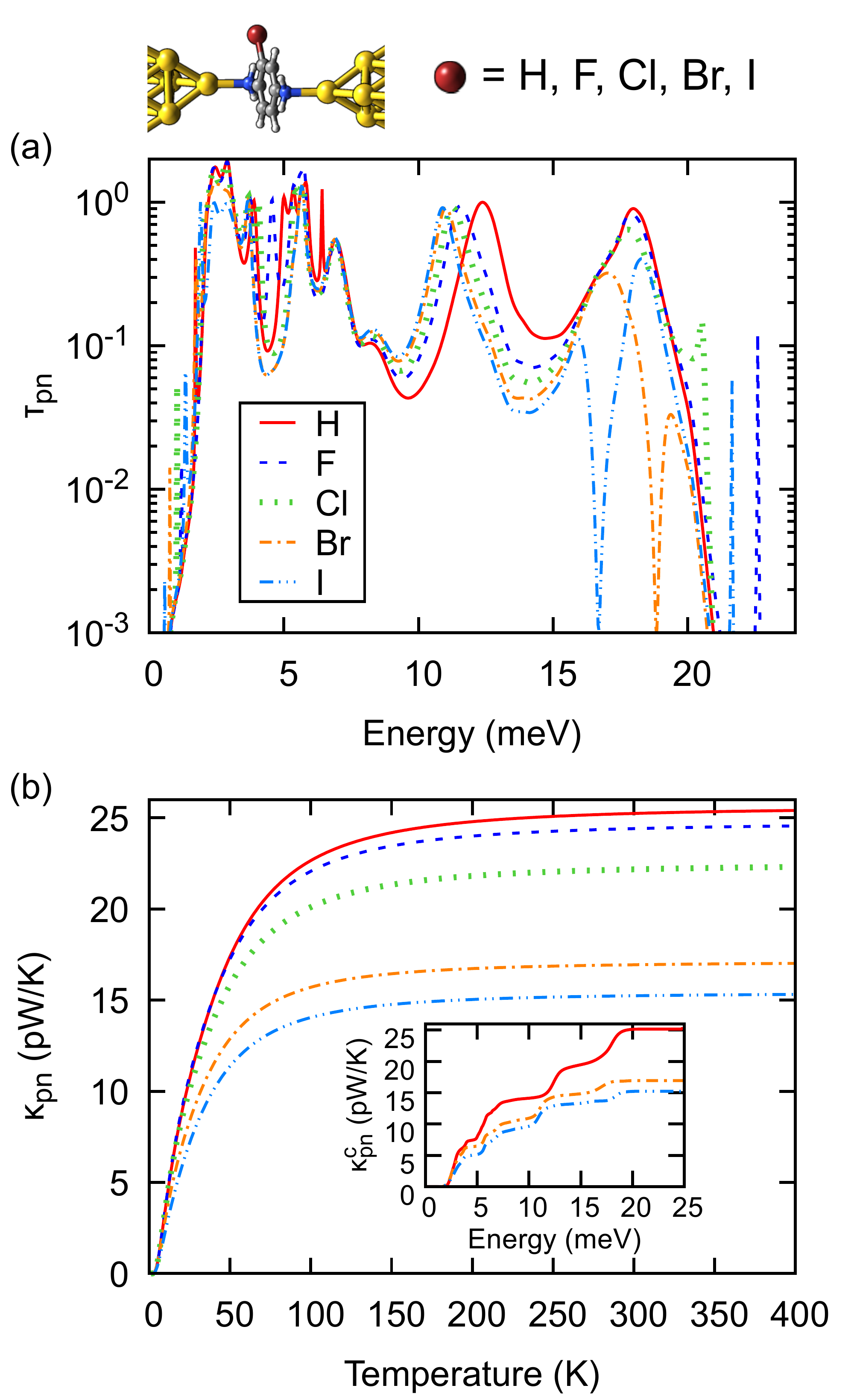}
\caption{(a) Phonon transmission as a function of energy for
  Au-benzenediamine-Au junctions, where a H atom of the benzene molecule has
  been substituted by a halogen atom ($X$=F, Cl, Br, I), see upper part of the
  figure. (b) The corresponding phononic thermal conductance as a function of
  temperature for the different benzenediamine
  derivatives. The inset shows the room-temperature
    cumulative thermal conductance as a function of energy for the junctions
    with $X$=H, Br, I. The legend in panel (a) applies also to panel (b) and
    its inset.}
  \label{fig-Benzene-para-subst}
\end{figure}

These results raise the question, whether it is possible to observe
interference effects in the phonon transport in junctions based on benzene
derivatives with the standard Au leads. In the context of electronic transport
it is known that interference features can be shifted in energy by introducing
side groups \cite{Andrews2008b}, which have either electron-withdrawing or
electron-donating character. The inclusion of such side groups moves the
resonance features to lower or higher energies, respectively. Furthermore, the
substituents can also break the symmetry of the molecule, leading to
destructive interference in benzene junctions even for the para configuration
of anchoring groups. Inspired by this idea, we shall analyze in what follows
the effect of substituting a H atom in the benzenediamine molecule by a
heavier atom of mass $m$ to tune the position of the resonance features. The
basic idea is to try to shift the destructive resonances below the Debye
energy of gold to observe a measurable effect on the thermal conductance.

With this idea in mind we consider the phonon transport in
Au-benzenediamine-Au single-molecule junctions, where one of the H atoms of
the benzene has been substituted by a halogen atom $X$=F, Cl, Br, I, as
depicted in the upper part of Fig.~\ref{fig-Benzene-para-subst}. The naive
expectation is that since the energy of a harmonic oscillator scales as $E
\propto \sqrt{k/m}$, with $k$ being the force constant, the resonance features
should decrease in energy with increasing mass $m$ of the substituent from F
to I. This simple view is indeed confirmed by our \emph{ab initio}
calculations of the phononic transmission, which are summarized in
Fig.~\ref{fig-Benzene-para-subst}(a) for the substituted benzenediamine
molecule in the para configuration.  As one can see, there is a clear
destructive interference feature for $X$=Br at an energy of around 19 meV. It
is further shifted to lower energies for $X$=I, where it appears at around 16
meV. Additionally, we see that the peak at around 13 meV for the unsubstituted
benzenediamine shifts to lower energies as the mass of the substituent
increases, while the transmission for energies lower than 10 meV remains
nearly unaffected. These results for the transmission have a clear impact on
the phononic thermal conductance, see Fig.~\ref{fig-Benzene-para-subst}(b).
Notice, in particular, that the thermal conductance decreases monotonically
with the mass of the substituent, reaching in the case of $X$=I a reduction
factor of 1.7 at room temperature, as compared with the unsubstituted benzene
molecule.

\begin{figure}[t]
\centering
\includegraphics[width=0.4\textwidth]{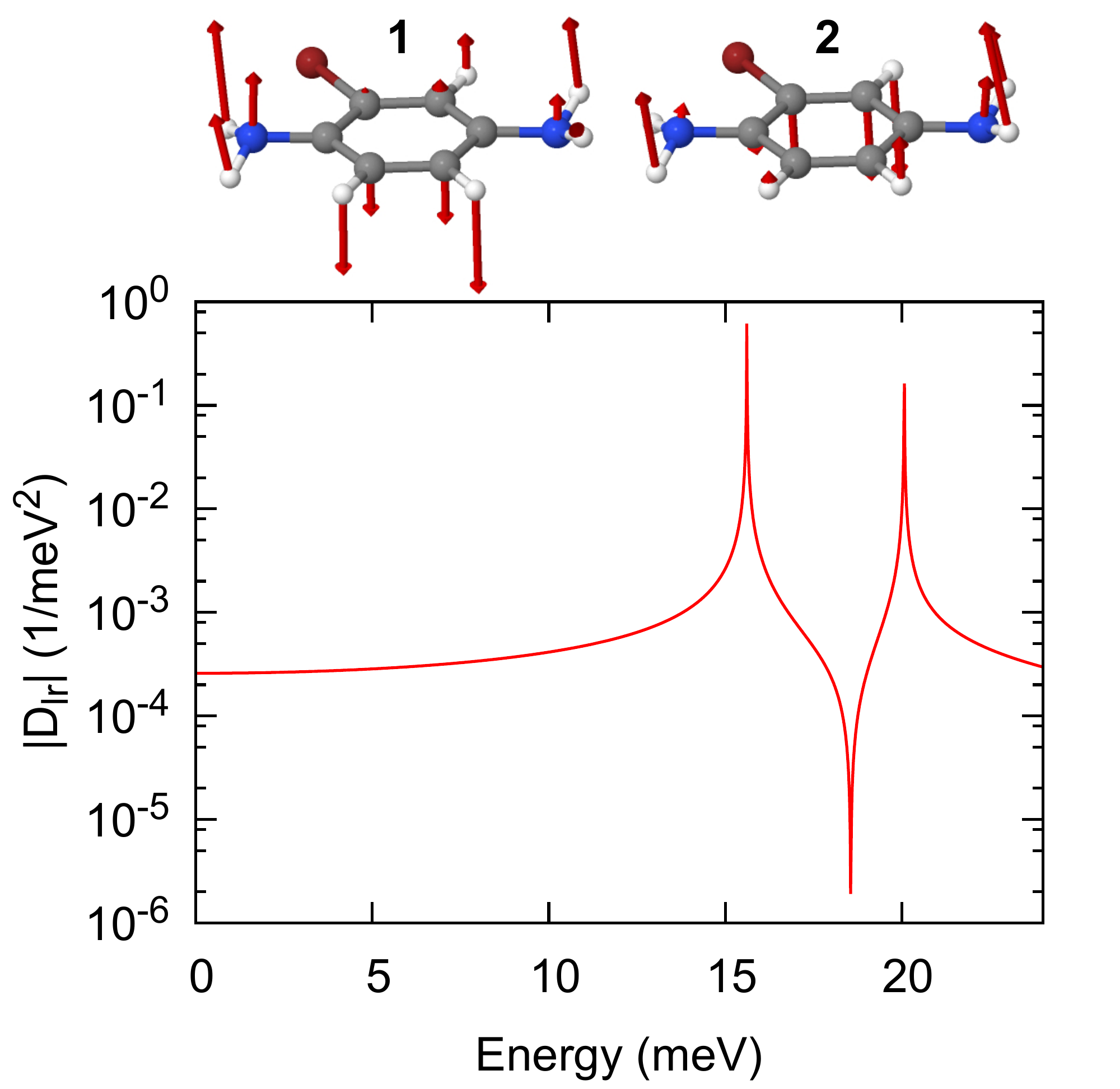}
\caption{In the upper part we visualize the two modes 1 and 2 that are
  responsible for the destructive interference feature at 19 meV in the
  benzenediamine molecule, substituted with $X$=Br. The first mode is at an
  energy of 15.61 meV and the second one at 20.07 meV. Since for both modes
  the arrows on the N atoms on the left and right sides point in the same
  direction, they interfere destructively. The graph below shows the
  calculated absolute value of the relevant Green's function element of the
  isolated molecule, taking into account in Eq.~(\ref{eq-Dlr}) only the modes
  1 and 2 and using $\eta=10^{-3}$~meV.}\label{fig-Dlr-Benzene}
\end{figure}
  To assess in a quantitative manner the impact of the
  antiresonances on the reduction of the thermal conductance upon the
  introduction of substituents, it is useful to investigate the so-called
  cumulative thermal conductance
\begin{equation}
  \kappa^{\rm c}_{\rm pn}(E,T) = \frac{1}{h} \int_{0}^{E} dE' \,E' \tau_{\rm pn}(E')
  \frac{\partial n(E',T)}{\partial T} \label{eq-kcph}
\end{equation}
 that is defined as the thermal conductance due to phonon modes up to a given
 energy $E$. We show in the inset of Fig.~\ref{fig-Benzene-para-subst}(b) the
 room-temperature cumulative thermal conductance as a function of energy for
 the Au-benzenediamine-Au junctions with $X$=H, Br, I. As one can see, the
 introduction of substituents actually modifies the phonon transport at all
 energies, but the change is particularly drastic at the energies at which the
 antiresonances occur. This illustrates that the destructive interference
 effects, induced by substituent atoms, play a key role in the reduction of
 the heat transport and can be used as a strategy to modify the thermal
 conductance.

In order to understand the origin of the destructive interference effect,
discussed above for Br and I, we follow an argument based on the symmetry of
the molecular orbitals developed for electron transport
\cite{Yoshizawa2008}. Since the structure of the transport formalism is almost
identical for electrons and phonons, with the only difference arising from the
equations of motion, this argument can be straightforwardly adapted to phonon
transport. Assuming for simplicity only nearest neighbor couplings, the idea
is as follows. If the molecule is attached to the left lead at position
$\text{l}$ with coupling constant $k_{\text{lL}}$ and at position $\text{r}$
with coupling constant $k_{\text{rR}}$ to the right lead, then the
transmission function can be expressed as
\begin{equation}\label{eq-tau_pn-approx}
  \tau_{\text{pn}}(E)=\frac{\pi^2}{E^2} k_{\text{lL}}^2
  k_{\text{rR}}^2\rho_{\text{L}}(E)\rho_{\text{R}}(E)\left|D_{\text{lr}}(E)\right|^2.
\end{equation}
Here $\rho_\alpha(E)$ is the local density of states at the lead atom
$\alpha=\text{L},\text{R}$ that is connected to the atom $\text{l},\text{r}$
of the molecule, respectively, and $D_{\text{lr}}(E)$ is the
$\text{lr}$-matrix element of the Green's function at energy $E$. Let us
assume that the typical embedding self-energies $\Pi_\alpha(E)$, which
describe the coupling of the central junction part to the left and right
electrodes\cite{Buerkle2015,Kloeckner2016} and appear in the full expression
for $D_{\text{lr}}(E)$, can be neglected, which is for instance the case if
molecule-lead couplings are not too strong. In this situation the Green's
function entering in the previous equation can be approximated by the
corresponding Green's function of the isolated molecule. It is given by the
following spectral representation
\begin{equation}\label{eq-Dlr}
  D_{\text{l}\text{r}}(E)=\sum_j \frac{C_{\text{l}j}C_{\text{r}j}^*}{\left(E+
    i\eta\right)^2 - E_j^2 },
\end{equation}
where $C_{nj}$ is the $n$-th component of the $j$-th eigenfunction or
vibrational mode. Given the dynamical matrix $\boldsymbol{K}$, the
eigenfunctions $C_{nj}$ and angular momentum frequencies $\omega_j$ are
obtained by solving the secular equation
\begin{equation}
  \boldsymbol{K} \boldsymbol{C}_j=\omega_j^2 \boldsymbol{C}_j.
\end{equation}
In the spectral representation in Eq.~(\ref{eq-Dlr}), $\eta$ is a small
imaginary part that prevents $D_{\text{lr}}(E)$ from diverging at
$E\rightarrow E_j$, and $E_j=\hbar \omega_j$ is the energy of the $j$-th
vibrational mode of the isolated molecule. As discussed in
Ref.~[\onlinecite{Yoshizawa2008}] for molecular junctions and in
Ref.~[\onlinecite{Lee1999}] for mesoscopic systems, a destructive interference
occurs between vibrational modes $j$ that exhibit the same sign of the product
$C_{\text{l}j}C_{\text{r}j}^*$, or more pictorially, the same parity of
vibrational modes at the lead-connecting molecular sites. This conclusion is
obvious from the general form of the Green's function in
Eq.~(\ref{eq-Dlr}). If the embedding self-energies
  $\Pi_\alpha(E)$ are taken into account, they may lead to renormalizations of
  level positions and hence antiresonance peaks. They are also important to
  describe level broadenings, which are mimicked in Eq.~(\ref{eq-Dlr}) by
  $\eta$, and hence to avoid any divergencies of the transmission in
  Eq.~(\ref{eq-tau_pn-approx}) at resonance positions $E=E_j$.

Let us now apply these ideas to understand the antiresonance that appears at
around 19 meV in the transmission function for the benzenediamine junction
with the substituent $X$=Br, see Fig.~\ref{fig-Benzene-para-subst}(a). For
this purpose we first analyzed the vibrational modes of the isolated molecule
within DFT and identified the two modes at 15.61 meV and 20.07 meV, which are
shown in the upper part of Fig.~\ref{fig-Dlr-Benzene}. Taking into account
only these two modes in the sum over $j$ in Eq.~(\ref{eq-Dlr}), we display in
this figure also the corresponding absolute value of $D_{\text{lr}}(E)$. Here
the connection to the leads has been assumed to be established only at the N
atoms and the coefficients $C_{nj}$ have been taken from the calculations of
the isolated molecule. Since these two modes exhibit the same
  oscillation direction on the N atoms, the products
  $C_{\text{l}1}C_{\text{r}1}^*$ and $C_{\text{l}2}C_{\text{r}2}^*$ have the
  same sign, and the modes 1 and 2 are expected to interfere destructively.
  This is confirmed by the behavior of $|D_{\text{lr}}(E)|$ in
  Fig.~\ref{fig-Dlr-Benzene}, which exhibits two peaks at the energies of the
  two modes and a minimum at around 19 meV, energetically located between
  these two modes. The position of the minimum is in good agreement with the
minimum in the transmission curve in Fig.~\ref{fig-Benzene-para-subst}. Let us
mention that the same considerations apply for the molecule with $X$=I, where
we find two modes of similar characteristics at energies around 14.7 and 19.5
meV.

One may wonder whether the antiresonances, resulting from
  destructive interference, survive upon the elongation of the molecular
  junctions. We have investigated this aspect for the junctions with the
  benzene derivatives studied in Fig.~\ref{fig-Benzene-para-subst} and found
  that the appearance of antiresonances is quite robust against stretching or
  compression. This is indeed expected since, as we have discussed above, the
  antiresonances originate from the interference of internal vibrational modes
  of the molecules, and they are therefore not greatly affected by variations
  of the distance between the electrodes.

\begin{figure}[t]
\centering
\includegraphics[width=0.4\textwidth]{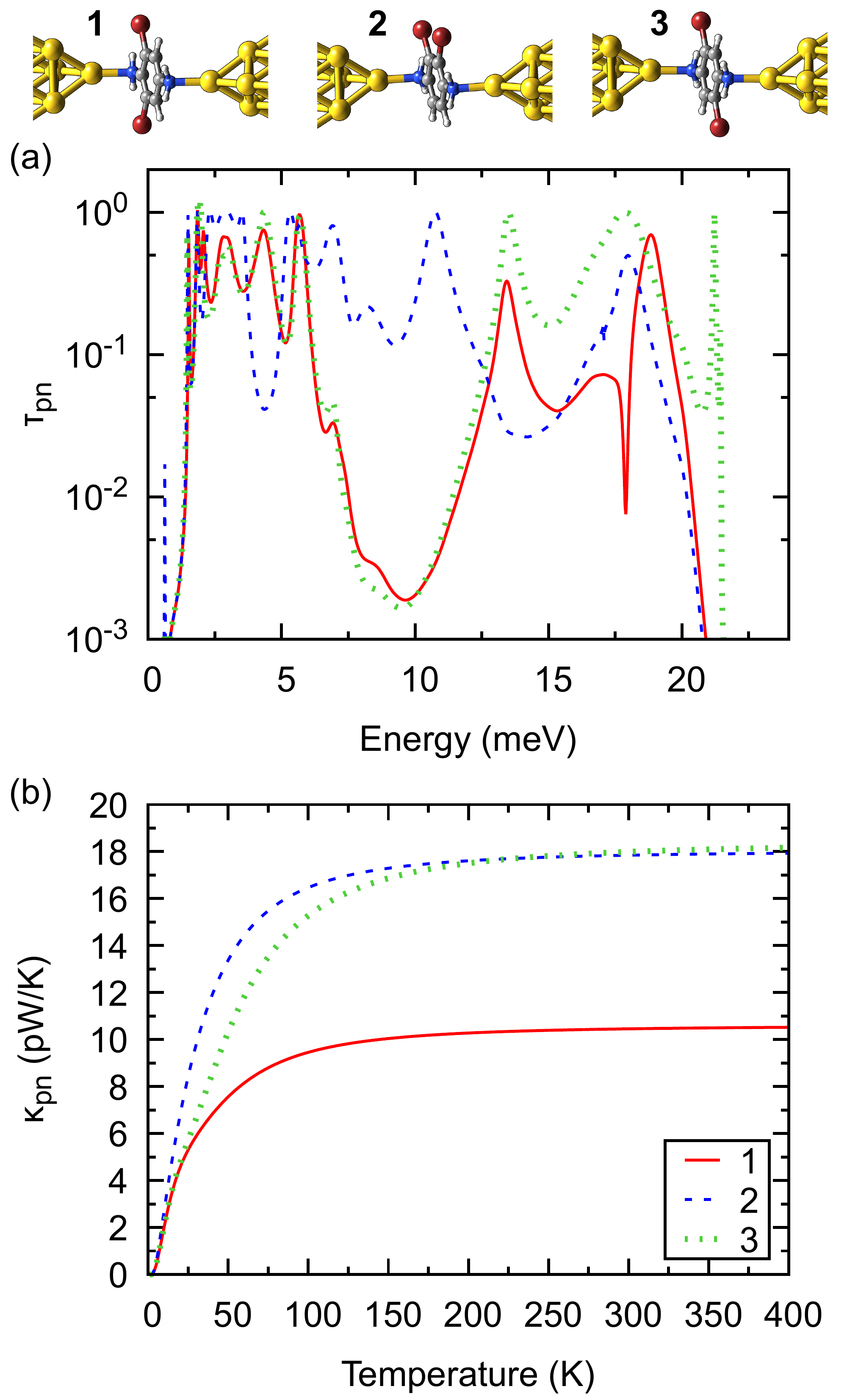}
\caption{(a) Phonon transmission as a function of energy for the three
  molecular junctions shown above this panel, where two H atoms of the
  benzenediamine molecule have been replaced by two Br atoms. (b) The
  corresponding temperature dependence of the phononic thermal
  conductance. The legend in panel (b) applies also to panel
    (a).}
\label{fig-Benzene-para-subst-2}
\end{figure}
\begin{figure}[t]
\centering
\includegraphics[width=\columnwidth]{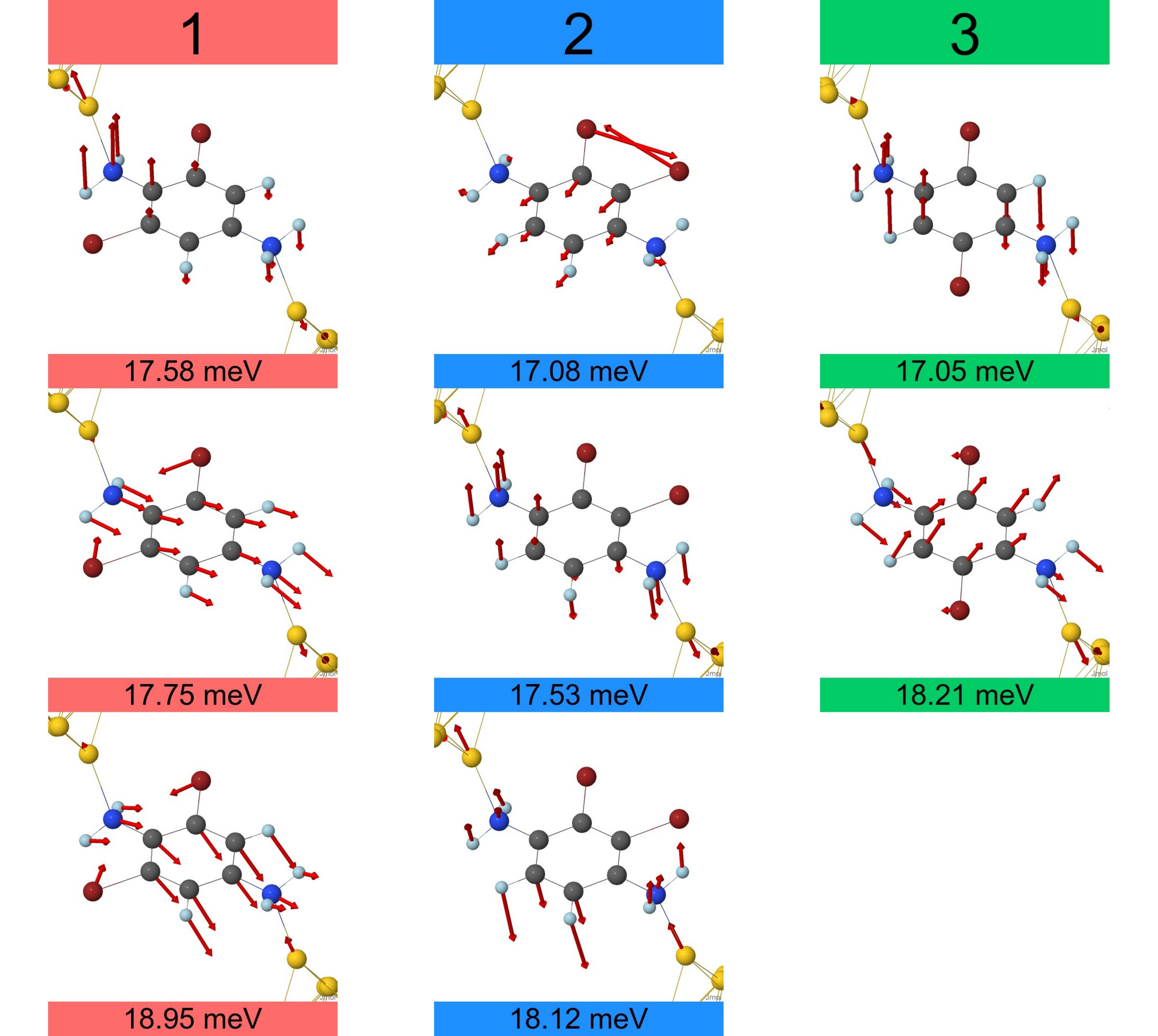}
\caption{All vibrational modes in the central region of the
    three molecular junctions discussed in Fig.~\ref{fig-Benzene-para-subst-2}
    in the energy range between 15 and 20 meV.}
\label{fig-modes-Br}
\end{figure}
 Following the analogy with electronic systems, we analyze now,
  how the inclusion of additional substituents modifies the interference
  patterns \cite{Garner2016}. For this purpose we investigate the phonon
  transport through benzenediamine junctions, if two H atoms are replaced by
  two Br atoms. Moreover we study the influence of the exact position, where
  these Br atoms are incorporated, and examine the three molecular junctions
  shown in the upper part of Fig.~\ref{fig-Benzene-para-subst-2}. All
  vibrational modes between 15 and 20 meV in the central region of the
  corresponding junctions are shown in Fig.~\ref{fig-modes-Br}.

The results for the phonon transmission are displayed in
  Fig.~\ref{fig-Benzene-para-subst-2}(a). As one can see, for molecule 1 a
  destructive interference antiresonance appears at about 17 meV. The origin
  of this feature is, as for the previous compounds, the interference between
  two modes that lie close in energy. These are the modes with energies of
  17.75 and 18.95 meV in the left column of Fig.~\ref{fig-modes-Br} that show
  the same parity on the terminal N atoms of the molecule. The difference with
  respect to the singly-substituted molecules discussed above is that these
  modes have no analogues in the isolated molecule. Instead they are hybrid
  modes in the sense that they also involve vibrations of the gold atoms in
  the electrodes. Additionally, a reduced transmission peak at 12.5 meV
  arises, which is due to a localized mode that is asymmetrically coupled to
  the electrodes. Interestingly, the other two molecules 2 and 3 do not
  exhibit any pronounced antiresonance, resulting from destructive
  interference, see Fig.~\ref{fig-Benzene-para-subst-2}(a), which we attribute
  to the lack of the necessary symmetry of the vibrational modes on the
  terminal N atoms, see Fig.~\ref{fig-modes-Br}.  This shows that not only
the masses of the substituents play a role, but also the exact position, where
they are introduced. The changes in the transmission are reflected in the
corresponding thermal conductance results. As we see in
Fig.~\ref{fig-Benzene-para-subst-2}(b), while molecule 1 exhibits a largely
reduced thermal conductance at room temperature as compared to both the singly
substituted case and the unsubstituted benzenediamine, molecules 2 and 3
exhibit conductance values that are similar to those of the singly substituted
case.

\begin{figure}[t]
\includegraphics[width=0.4\textwidth]{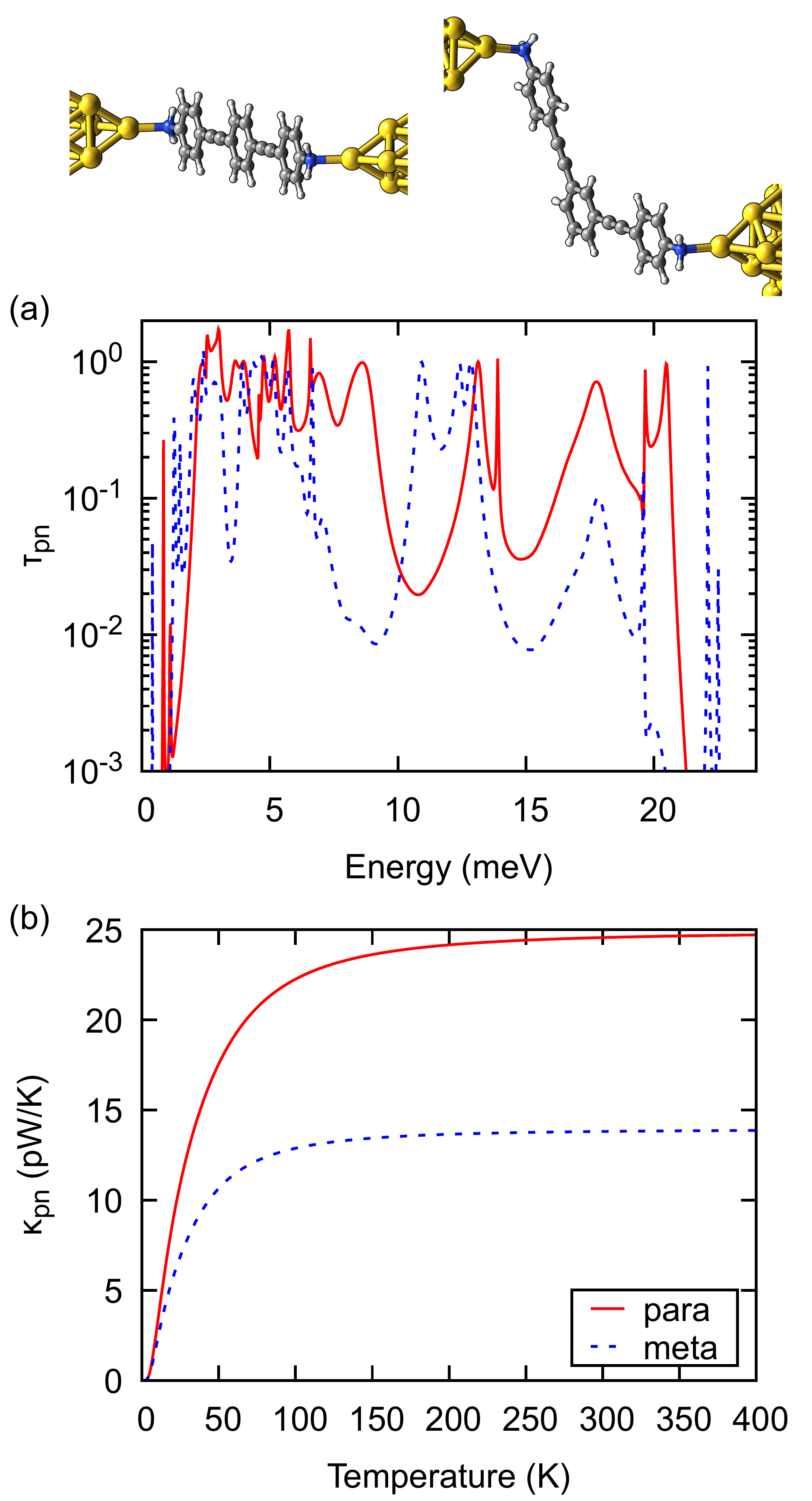}
\caption{(a) Phonon transmission as a function of energy for the Au-OPE3-Au
  junctions shown above this panel for both para (left) and meta (right)
  binding configurations at the central benzene ring. (b) The corresponding
  phononic thermal conductance as a function of
  temperature. The legend in panel (b) applies also to panel
    (a).}
\label{fig-OPE3-para-meta}
\end{figure}
Now we show that the basic concepts discussed above also apply to other, more
complex molecules. For this purpose, we consider OPE3. This molecule has been
analyzed in the context of phonon transport in Ref.~[\onlinecite{Markussen2013}]
with the help of semi-empirical methods and considering Si as well as
graphene-nanoribbon electrodes. First we discuss the influence of the binding
configuration, by examining para- vs.\ meta-OPE3, as shown in the upper part
of Fig.~\ref{fig-OPE3-para-meta}.  In this case our \emph{ab initio} results
for the phonon transmission, which are displayed in
Fig.~\ref{fig-OPE3-para-meta}(a), show that for meta-OPE3 several resonance
peaks at energies below 15 meV appear shifted to lower energies as compared to
para-OPE3, while there is a pronounced decrease of the transmission for the
meta-OPE3 above. The strong suppression of the transmission at around 18 meV
is due to an interference effect, in which two quasi-degenerate modes are
involved, as explained in Ref.~[\onlinecite{Markussen2013}]. In total the
meta-OPE3 thermal conductance is reduced by a factor of 2, as compared to
para-OPE3.

\begin{figure}[t]
\includegraphics[width=0.4\textwidth]{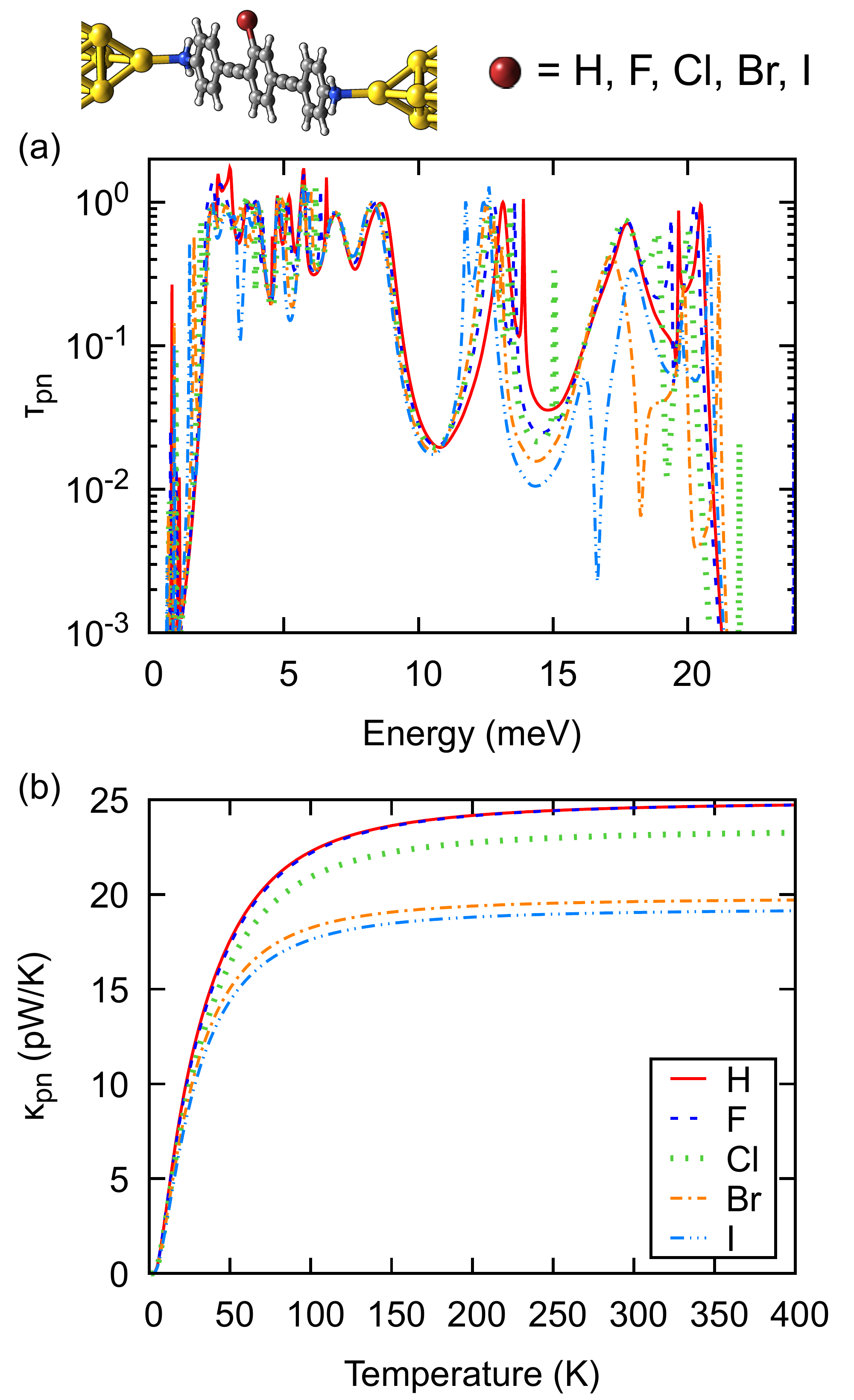}
\caption{(a) Phonon transmission as a function of energy for Au-para-OPE3-Au
  junctions, where a H atom of the central benzene ring has been substituted
  by a halogen atom ($X$=F, Cl, Br, I), see plot above this panel. (b) The
  corresponding phononic thermal conductance as a function of temperature for
  the different para-OPE3 derivatives. The legend in panel
    (b) applies also to panel (a).}
\label{fig-OPE3-para-subst}
\end{figure}

As in the case of benzene, the thermal conductance of Au-OPE3-Au junctions can
be tuned by substituting one H atom in the central benzene ring with a halogen
atom, as sketched in the upper part of Fig.~\ref{fig-OPE3-para-subst}. Similar
to benzenediamine in Fig.~\ref{fig-Benzene-para-subst}, one can see in
Fig.~\ref{fig-OPE3-para-subst}(a) that the phonon transmission exhibits an
antiresonance at energies around 15-20 meV for the substituents Br and I, but
now this feature also appears for Cl. Again the occurrence of destructive
interferences below the Debye energy of gold leads to a suppression of the
corresponding thermal conductance, as we show in
Fig.~\ref{fig-OPE3-para-subst}(b). We find a monotonically decreasing thermal
conductance with increasing mass of the substituent. As in the
case of the benzene derivatives this behavior is due to the fact that, upon
introducing the substituents, the energies of the vibrational modes decrease
and the antiresonances, resulting from the interference effects, are
redshifted at the same time.  Thus, since the contribution of a vibrational
mode scales with its energy, see Eq.~(\ref{eq-kph}), $\kappa_{\rm pn}$ is
reduced upon increasing mass of the substituents.
 
Let us mention that we are not aware that the exact OPE3 compounds, 
shown in Fig.~\ref{fig-OPE3-para-subst}, have been synthesized. But closely related 
molecules exist, in which the central ring of an OPE3-diamine has been modified, 
among others, by the inclusion of several fluorine atoms and which have been 
studied in the context of single-molecule junction experiments \cite{Gonzalez2014}. 
So, we think that there should not be any fundamental problem to synthesize the 
compounds discussed in this work. In this respect, it is also worth stressing 
that our DFT calculations demonstrate that these molecules are indeed stable
  
To conclude the discussion of our results on phonon heat transport, it is important 
to emphasize that in all the molecular junctions investigated in this work, the 
room-temperature thermal conductance is dominated by the contribution of phonons. 
Therefore the impact of the interference effects discussed here should be, in principle, 
amenable to measurement. This point is discussed in detail in Appendix \ref{sec-elthcond}, 
where we present our results for the electronic contribution to the thermal conductance.

\section{Conclusions}

Making use of a full \emph{ab initio} transport method, we have analyzed the
influence of phonon interference on the thermal conductance of benzene and
OPE3 derivatives attached to Au electrodes via amino groups. We have found
that for unsubstituted benzene no interference effects are visible,
irrespective of the binding configuration, which is due to the small Debye
energy of gold. We have also shown that by substituting one H atom with a
halogen atom of increasing mass leads to the appearance of Fano-like
resonances at energies below the Au Debye energy, which in turn leads to a
reduction of thermal conductances by up to a factor of 1.7. We were able to
relate this kind of antiresonance feature to the destructive interference
between two modes of the free molecule with the same symmetry in the region,
where the molecule is connected to the leads. We have also shown that the
thermal conductance can be further reduced by increasing the number of
substituent atoms and arranging them in appropriate positions.

Finally, we have used similar concepts to tune the thermal conductance of
junctions based on OPE3 molecules. In particular, we have demonstrated that
the thermal conductance of meta-OPE3 junctions is clearly reduced as compared
to the para-OPE3 case mainly due to an interference between two
quasi-degenerate vibrational modes. On the other hand, we have also shown that
the thermal conductance of a linear Au-OPE3-Au junction can be
reduced by incorporating halogen atoms in the central ring of
this molecule. Again, this effect is due to the appearance of destructive
phonon interference.

In summary, we have demonstrated that, in analogy to the electronic transport,
the thermal conductance of molecular junctions can be controlled by means of
phonon interference effects. Even if the effect for phonons is not as dramatic
as for the electronic analog, our findings are important to achieve
heat management in nanostructured metal-molecule hybrid systems and to explore
their potential in thermoelectric applications.

\section{Acknowledgment}

J.C.K.\ thanks A.\ Irmler for stimulating discussions. J.C.K.\ and
F.P.\ gratefully acknowledge funding from the Carl Zeiss foundation and the
Junior Professorship Program of the Ministry of Science, Research, and the
Arts of the state of Baden W\"urttemberg. J.C.C.\ is supported by the German
Research Foundation (DFG) and the Collaborative Research Center (SFB) 767,
which sponsor his stay at the University of Konstanz as Mercator Fellow, as
well as the Spanish Ministry of Economy and Competitiveness (Contract
No.\ FIS2014-53488-P). An important part of the numerical modeling was carried
out on the computational resources of the bwHPC program, namely the
bwUniCluster and the JUSTUS HPC facility.

\appendix

\section{Electronic thermal conductance}
\label{sec-elthcond}

In this appendix we briefly discuss our results on the
  electronic contribution to the thermal conductance for the different
  molecular junctions investigated in this work. The goal is to show that in
  these junctions the thermal transport is dominated by phonons and that the
  interference effects, predicted here, should therefore be observable
  experimentally. We neglect contributions to the thermal conductance from
  near field radiative heat transfer, whose significance can be controlled by
  the macroscopic shape of the electrodes \cite{Kloeckner2017}.
  
  As in the phononic case we assume that the electronic
  transport is dominated by elastic tunneling processes. Thus, we compute the
  electronic contribution to the thermal conductance in the linear response
  regime within the Landauer-B\"uttiker formalism. It is given by
  \cite{Cuevas2017}
  \begin{equation}
    \kappa_{\rm el}(T) = \frac{2}{hT} \left( K_{2}(T)- \frac{K_{1}(T)^{2}}{K_{0}(T)}
    \right), \label{eq-kel}
  \end{equation}
  where the coefficients are defined as
  \begin{equation}
    K_n(T) = \int^{\infty}_{-\infty} dE \, \tau_{\rm el}(E) \left(-\frac{\partial
      f(E,T)}{\partial E}\right)(E-\mu)^{n},
    \label{eq-Kn}
  \end{equation}
  $\tau_{\rm el}(E)$ is the energy-dependent electron transmission, and
  $f(E,T) = \left\{ \exp[(E-\mu)/k_{\mathrm{B}}T]+1\right\} ^{-1}$ is the
  Fermi function. Here, the chemical potential $\mu\approx E_{\rm F}$
  is approximately given by the Fermi energy $E_{\rm F}$ of the Au
  electrodes.
  
  To compute the electronic transmission function, we have
  employed an approach based on the combination of DFT and nonequilibrium
  Green's function techniques that we have interfaced to TURBOMOLE and
  explained in detail in Ref.~[\onlinecite{Pauly2008}]. Moreover, in order to
  correct for the known inaccuracies in DFT related to quasiparticle energies,
  we have made use of the DFT+$\Sigma$ approach \cite{Quek2007}, which was
  implemented in our quantum transport method as described in
  Ref.~[\onlinecite{Zotti2014}].

\begin{figure}[t]
\includegraphics[width=\columnwidth]{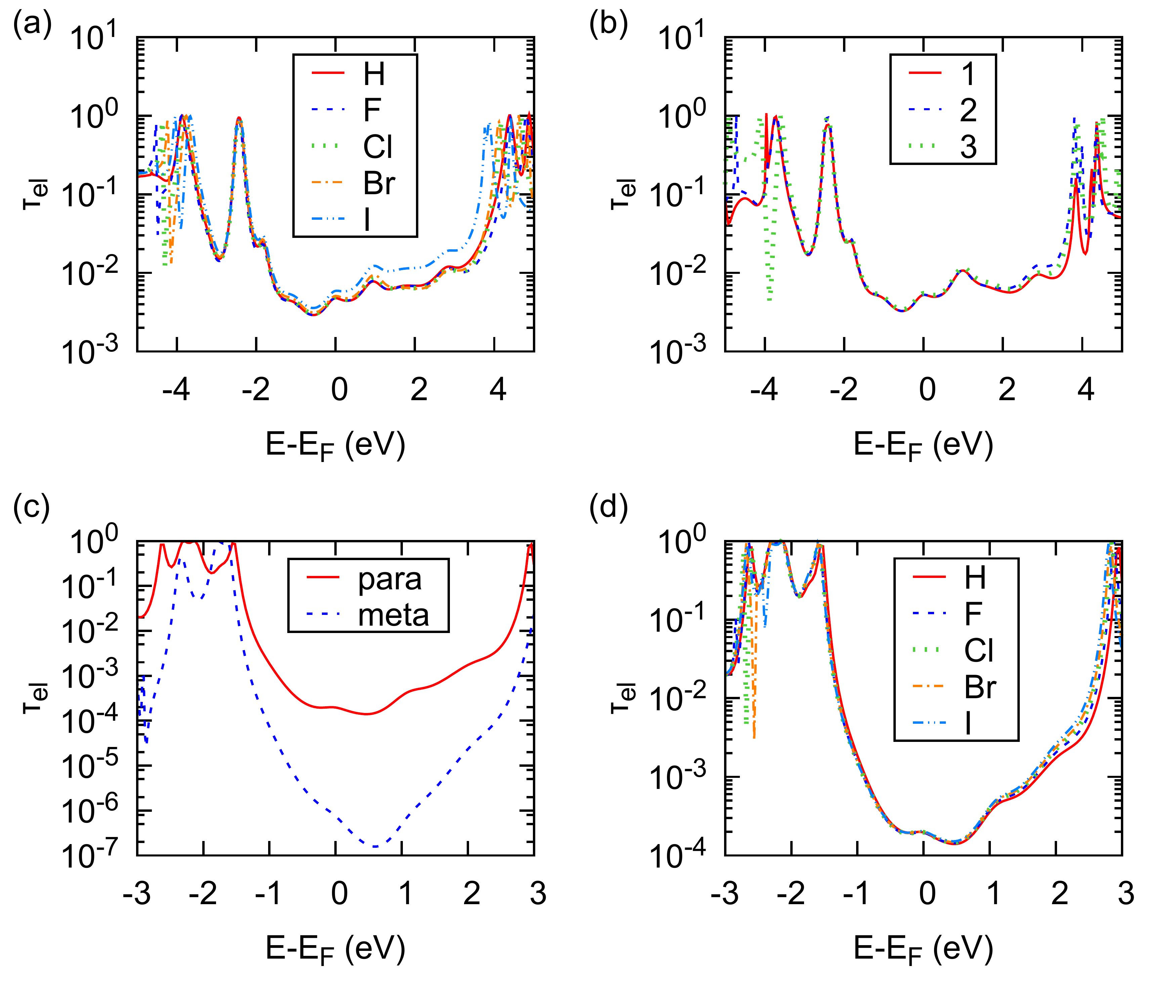}
\caption{Electronic transmission as a function of energy,
    computed for (a) the molecular junctions of
    Fig.~\ref{fig-Benzene-para-subst} based on benzene with a single halogen
    atom as substituent, (b) the junctions of
    Fig.~\ref{fig-Benzene-para-subst-2} based on benzene with two Br atoms as
    substituents, (c) the para- and meta-bonded OPE3 junctions of
    Fig.~\ref{fig-OPE3-para-meta} and (d) the OPE3-based junctions of
    Fig.~\ref{fig-OPE3-para-subst} with a single substituent on the central
    benzene ring.}
\label{fig-tauel}
\end{figure}
\begin{figure}[b]
  \includegraphics[width=\columnwidth]{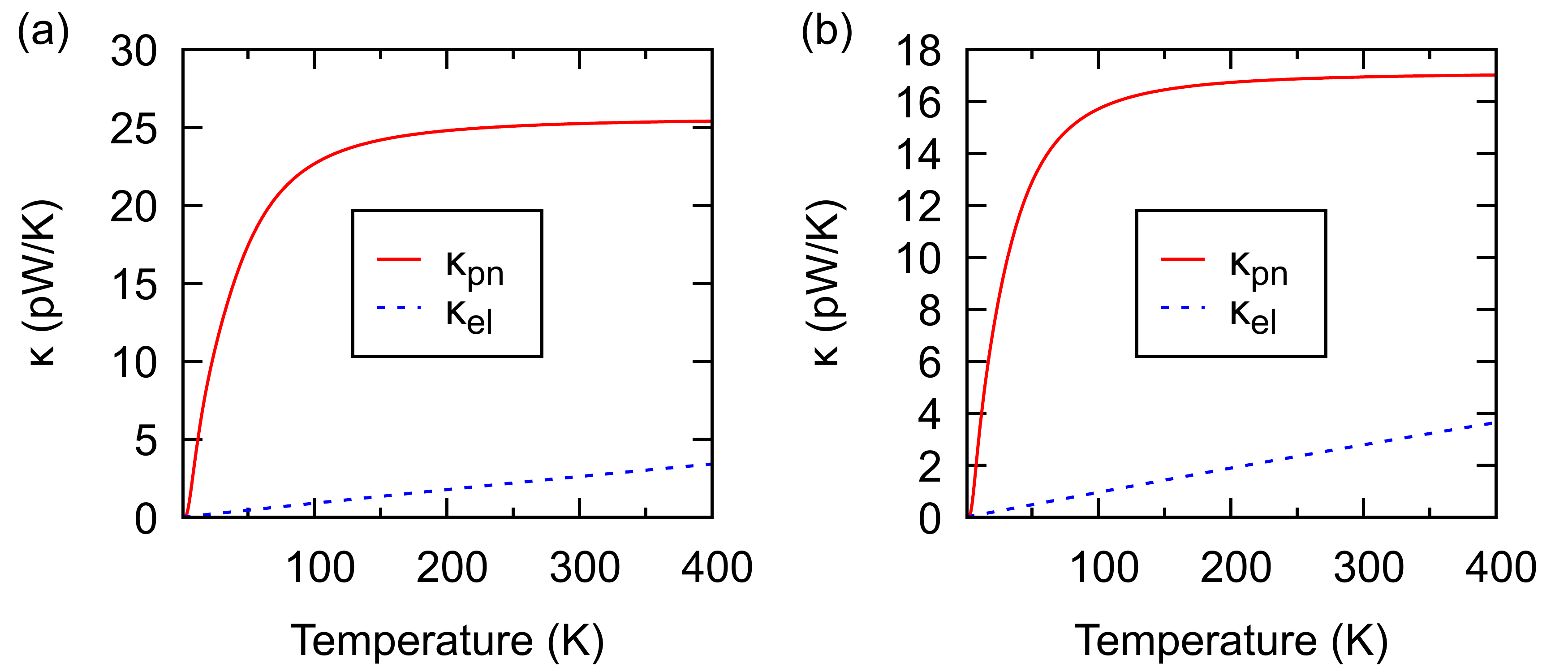}
  \caption{Temperature dependence of the phononic and
      electronic contribution to the thermal conductance for the (a)
      Au-1,4-diaminobenzene-Au and (b) Au-2-bromo-1,4-diaminobenzene-Au
      junctions of Fig.~\ref{fig-Benzene-para-subst}.}
  \label{fig-comp}
\end{figure}

  In Fig.~\ref{fig-tauel} we show our results for the electronic
  transmission $\tau_{\rm el}$ as a function of energy for all the junctions
  of Figs.~\ref{fig-Benzene-para-subst}, \ref{fig-Benzene-para-subst-2},
  \ref{fig-OPE3-para-meta}, and \ref{fig-OPE3-para-subst} based on both
  benzene and OPE3 derivatives. In all cases the electronic transport proceeds
  mainly through the tail of the HOMO of the molecule in an off-resonant
  situation.  The results for the transport properties of these molecular
  junctions at room temperature are summarized in Table~\ref{table}. We
  present there both the phononic and electronic thermal conductances
  $\kappa_{\rm pn}$ and $\kappa_{\rm el}$, as well as the electrical
  conductance
  \begin{equation}
    G(T)=G_0K_0(T),
  \end{equation}
  with $G_0=2e^2/h$. Moreover, when available, we also report the experimental
  values for $G$. The electric conductance can be used together with the
  Wiedemann-Franz law $\kappa_{el}\approx L_0GT$ with the Lorentz number
  $L_0=(k_{\rm B}/e)^2\pi^2/3$ to estimate the electronic thermal
  conductance. We find the Wiedemann-Franz law to be approximately fulfilled
  for our computed molecular junctions, and the comparison of experimental and
  theoretical electrical conductance values $G$ can hence be used to estimate
  uncertainties in the theoretical $\kappa_{\rm el}$.
  
  The key point is that in all the studied cases the
    electronic thermal conductance is considerably smaller than the
    corresponding phononic thermal conductance. In particular, for the
    junctions based on the benzene derivatives the electronic contribution is
    at least 5 times smaller than the phononic one, while for the OPE3
    compounds the electronic contribution is more than two orders of magnitude
    smaller. Due to their lower $G$ the longer molecules are thus
    advantageous, if we want to exclude the contribution of electrons to heat
    transport.

Moreover, in order to illustrate that phonons dominate the
  thermal transport in these junctions for a wide temperature range, we show
  in Fig.~\ref{fig-comp} the temperature dependence of the phononic and
  electronic thermal conductances for the Au-1,4-diaminobenzene-Au and
  Au-2-bromo-1,4-diaminobenzene-Au junctions of
  Fig.~\ref{fig-Benzene-para-subst}. As one can see, the phononic contribution
  largely dominates at all temperatures between 10 and 400 K. Similar results
  hold for all the other molecular junctions.

To conclude, the results presented in this appendix confirm
  that the thermal transport in the molecular junctions studied in this work
  is dominated by phonons. The predicted interference effects should therefore
  be visible in possible experiments.

\begin{widetext}
\begin{center}
\begin{table}[t]
\caption{Computed room-temperature phononic thermal
    conductance $\kappa_{\rm pn}$, electronic thermal conductance $\kappa_{\rm
      el}$ and electrical conductance $G$ in units of the electrical
    conductance quantum $G_0=2e^2/h$ for the different molecular junctions
    investigated in this work. The last column shows, when available, the
    experimental value of the electrical conductance, as obtained from the
    peaks of conductance histograms. Experimental uncertainties due to broad
    distributions of conductance values have been omitted. \label{table}}
\begin{tabular}{lcccc}
\hline 
{\bf Molecule} & $\kappa_{\rm pn}$ (pW/K) & $\kappa_{\rm el}$ (pW/K) & 
$G$ ($G_0$) (theory) & $G$ ($G_0$) (exp.) \tabularnewline
\hline 
1,4-diaminobenzene (para) & 25.24 & 2.61 & $4.7 \times 10^{-3}$ & $6.4 \times 10^{-3}$ 
[\onlinecite{Venkataraman2007}] \tabularnewline
2-fluoro-1,4-diaminobenzene & 24.40 & 2.62 & $4.7 \times 10^{-3}$ &
$5.8 \times 10^{-3}$ [\onlinecite{Venkataraman2007}] \tabularnewline
2-chloro-1,4-diaminobenzene & 22.17 & 2.70 & $4.9 \times 10^{-3}$ & 
$6.0 \times 10^{-3}$ [\onlinecite{Venkataraman2007}] \tabularnewline
2-bromo-1,4-diaminobenzene & 16.94 & 2.78 & $5.0 \times 10^{-3}$ &
$6.1 \times 10^{-3}$ [\onlinecite{Venkataraman2007}] \tabularnewline
2-iodo-1,4-diaminobenzene & 15.24 & 3.25 & $5.8 \times 10^{-3}$ &
\tabularnewline
2,5-dibromo-1,4-diaminobenzene & 17.85 & 2.88 & $5.20 \times 10^{-3}$
\tabularnewline
2,6-dibromo-1,4-diaminobenzene & 10.46 & 2.88 & $5.19 \times 10^{-3}$ 
\tabularnewline
2,3-dibromo-1,4-diaminobenzene & 17.99 & 3.02 & $5.44 \times 10^{-3}$ 
\tabularnewline
OPE3 (para) & 24.57 & 0.11 & $1.94 \times 10^{-4}$ & $2.6 \times 10^{-5}$ 
[\onlinecite{Gonzalez2013}], $1.27 \times 10^{-4}$ [\onlinecite{Lu2009}] \tabularnewline
OPE3 (meta) & 13.82 & $4.24 \times 10^{-4}$ & $7.53 \times 10^{-7}$ & \tabularnewline
F-OPE3 & 24.56 & 0.11 & $1.93 \times 10^{-4}$ & \tabularnewline
Cl-OPE3 & 23.12 & 0.11 & $1.96\times 10^{-4}$ & \tabularnewline
Br-OPE3 & 19.62 & 0.11 & $1.97\times 10^{-4}$ & \tabularnewline
I-OPE3 & 19.04 & 0.11 & $2.00\times 10^{-4}$ & \tabularnewline

\hline 
\end{tabular}
\end{table}
\end{center}
\end{widetext}

\bibliographystyle{apsrev4-1} 

\end{document}